\documentclass[12pt,preprint]{aastex}
\usepackage{graphicx,subfigure,amsmath,hyperref}
\newcommand\gsim{\,\lower3pt\hbox{$\sim$}\llap{\raise2pt\hbox{$>$}}\,}
\newcommand\lsim{\,\lower3pt\hbox{$\sim$}\llap{\raise2pt\hbox{$<$}}\,}

\begin{document}
\title{Thermal signatures of tether-cutting reconnections in pre-eruption
coronal flux ropes: hot central voids in coronal cavities}

\author{Y.~Fan}
\affil{High Altitude Observatory, National Center for Atmospheric
Research\altaffilmark{1}, 3080 Center Green Drive, Boulder, CO 80301}

\altaffiltext{1} {The National Center for Atmospheric Research
is sponsored by the National Science Foundation}

\begin{abstract}
Using a 3D MHD simulation, we model the quasi-static evolution and the onset of
eruption of a coronal flux rope. The simulation begins with a twisted flux
rope emerging at the lower boundary and pushing into a pre-existing
coronal potential arcade field. At a chosen time the emergence is stopped
with the lower boundary taken to be rigid. Then the coronal flux rope settles
into a quasi-static rise phase during which an underlying, central
sigmoid-shaped current layer forms along the so called hyperbolic flux
tube (HFT), a generalization of the X-line configuration.
Reconnections in the dissipating current layer effectively add twisted flux
to the flux rope and thus allow it to rise quasi-statically, even though the
magnetic energy is decreasing as the system relaxes.
We examine the thermal features produced by the current layer formation and
the associated ``tether-cutting''
reconnections as a result of heating and field aligned thermal conduction.
It is found that a central hot,
low-density channel containing reconnected,
twisted flux threading under the flux rope axis forms on top of the central
current layer. When viewed in the line of sight roughly aligned with the hot
channel (which is roughly along the neutral line), the central current layer
appears as a high-density vertical column with upward extensions as a ``U''
shaped dense shell enclosing a central hot,
low-density void. Such thermal features have
been observed within coronal prominence cavities. Our MHD simulation suggests
that they are the signatures of the development of the HFT topology and the
associated tether-cutting reconnections, and that the central void grows
and rises with the reconnections, until the flux rope reaches the critical
height for the onset of the torus instability and dynamic eruption ensues.
\end{abstract}

\section{Introduction}
CMEs and eruptive flares are believed to result from a sudden, explosive
release of the free magnetic energy stored in the previously
quasi-equilibrium, twisted/sheared coronal magnetic
field \citep[see e.g. reviews by][]{forbesetal_2006,chen2011}.
Many CME models have considered a magnetic flux rope containing helical field
lines twisting about each other in the corona as the basic underlying magnetic
field structure for CME precursors \citep[e.g.][]{td1999,low2001,
gibson_fan2006}.
Recent observational studies of coronal prominence cavities support
the picture of twisted coronal flux ropes as CME precursors \citep[e.g.][]{wang_stenborg2010}. Theoretical studies have shown that a coronal flux rope
confined by an external potential magnetic field can become unstable to
the torus instability and erupt when it reaches a critical height where the
ambient potential field decreases with height too steeply
\citep{kliem_toeroek2006,isen_forbes2007,demoulin_aulanier2010}.
MHD simulations have been carried out to study the critical condition and
the non-linear evolution of the torus instability of various 3D coronal flux
rope structures \citep[e.g.][]{toeroek_kliem2007,fan_gibson2007,
aulanieretal2010,fan2010}.

Through a sequence of 3D MHD simulations of the evolution of the coronal
magnetic field resulting from the emergence of a twisted magnetic flux rope
into a pre-existing coronal arcade field, \citet{fan2010} studied the
conditions that lead to a dynamic eruption of the resulting coronal flux rope.
It is found that the critical condition for the onset of eruption is for the
center of the flux rope to reach a critical height at which the corresponding
potential field declines with height at a sufficiently steep rate, consistent
with the onset of the torus instability of the flux rope. The simulations
show that after the flux emergence is stopped, the coronal flux rope first
settles into a quasi-static rise phase with an underlying, central
sigmoid-shaped current layer developing. Reconnections in the current layer
is found to effectively add twisted flux to the flux rope
\citep[as quantified by eq. (14) in][]{fan2010}, allowing it to rise
quasi-statically, even as the magnetic energy is declining.
We thus call these reconnections ``tether-cutting" reconnections in the
sense that they are effectively reducing the anchoring of the flux rope by
dissipating the underlying, central current layer whose current attracts
the volume current in the flux rope \citep{low_zhang_2002}.
When the flux rope rises quasi-statically to the critical height for the
onset of the torus instability, then dynamic eruption ensues.
To identify the thermal signatures that may result from the 
formation of the current layer and the associated reconnections, we carry
out an MHD simulation with a similar setup as those in \citet{fan2010}, but
with a more sophisticated treatment of the thermodynamics.
Instead of assuming a simple isothermal evolution, we solve the total energy
equation in conservative form for a perfect gas with an adiabatic index
$\gamma = 1.1$. In this way the
non-adiabatic heating due to the dissipation of kinetic and magnetic
energies by the numerical diffusions are implicitly incorporated in the
thermal energy. Also thermal conduction along the magnetic field lines is
included in the energy equation.
We find from this new simulation that the main consequence of the
tether-cutting reconnections in the current layers during the quasi-static
phase is the formation and enlargement of a hot, low density channel,
containing reconnected, twisted flux threading under the axis of the
flux rope. When viewed above the limb along a line of sight that is roughly
aligned with the channel, the thermal signature would manifest in SDO/AIA
observations as a hot central void 
enclosed in a ``U'' shaped dense shell on top of a relatively dense structure
inside the flux rope.
This signature inside the flux rope is consistent with the recent
observations of the thermal sub-structures inside coronal prominence cavities
by \citet{reevesetal2012, berger2012, regnieretal2011}.

We organize the remainder of the paper as follows. In Section \ref{sec:model},
we describe the MHD numerical model with the more sophisticated treatment
of the thermodynamics compared to the previous simulations in \citet{fan2010}.
In Section \ref{sec:result} we describe the simulation result, focusing on
the thermal signature that develops inside the flux rope and compare it
with observations.  We summarize and discuss the conclusions in
Section \ref{sec:conc}.

\section{Model Description\label{sec:model}}
For the simulation carried out in this study, we solve the following
magneto-hydrodynamic equations in a spherical domain:
\begin{equation}
\frac{\partial \rho}{\partial t}
+ \nabla \cdot ( \rho {\bf v}) = 0 ,
\label{eqcont}
\end{equation}
\begin{equation}
\rho \left ( \frac{\partial {\bf v}}{dt}
+ ({\bf v} \cdot \nabla ) {\bf v} \right )
= - \nabla p - \rho \frac{G M_{\odot}}{r^2} \hat{\bf r}+ \frac{1}{4 \pi}
( \nabla \times {\bf B} ) \times {\bf B},
\label{eqmotion}
\end{equation}
\begin{equation}
\frac{\partial {\bf B}}{\partial t}
= \nabla \times ({\bf v} \times {\bf B}),
\label{eqinduc}
\end{equation}
\begin{equation}
\nabla \cdot {\bf B} = 0,
\label{eqdivb}
\end{equation}
\begin{equation}
\frac{\partial e}{\partial t} = - \nabla \cdot
\left [ \left ( \varepsilon + \rho \frac{v^2}{2} + p \right ) {\bf v}
- \frac{1}{4 \pi} ({\bf v} \times {\bf B} ) \times {\bf B} \right ]
- \rho {\bf v} \cdot \frac{G M_{\odot}}{r^2} \hat{\bf r} - \nabla \cdot
{\bf q}  ,
\label{eqetot}
\end{equation}
\begin{equation}
p = \frac{\rho R T}{\mu},
\label{eqstate}
\end{equation}
where
\begin{equation}
\varepsilon = {\frac{p}{\gamma - 1} } .
\end{equation}
\begin{equation}
e=\varepsilon + \rho \frac{v^2}{2} + \frac{B^2}{8 \pi} .
\end{equation}
\begin{equation}
{\bf q} = -\kappa_0 T^{5/2} \hat{\bf b} \hat{\bf b} \cdot \nabla T
\label{conductflux}
\end{equation}
In the above ${\bf v}$ is the velocity field, ${\bf B}$ is the magnetic field,
$\rho$, $p$, and $T$ are respectively the plasma density, pressure and temperature,
$\varepsilon$, $e$ $R$, $\mu$, and $\gamma$, are respectively
the internal energy density, the total energy density (internal+kinetic+magnetic),
the gas constant, the mean molecular weight, and the adiabatic index of
the perfect gas,
${\bf q}$ is the heat flux due to field aligned thermal conduction, with the field direction
denoted by the unit vector ${\hat {\bf b}}$ and with $\kappa_0 = 10^{-6} \,
{\rm erg} \, {\rm K}^{-7/2} \, {\rm cm}^{-1} \, {\rm s}^{-1}$,
and $G$ and $M_{\odot}$ denote the gravitational constant and the mass
of the Sun.  We have assumed a perfect gas with a low value of
the adiabatic index: $\gamma = 1.1$, which allows the coronal plasma
to better maintain its (high) temperature during adiabatic expansion.
No {\it explicit} viscosity and magnetic diffusion are included in the
momentum and the induction equations.
However numerical dissipations are present at regions of sharp gradient, and
since we are solving the total energy equation of the plasma in conservative
form (eq. [\ref{eqetot}]), the non-adiabatic heating due numerical dissipation
of kinetic, and magnetic energies is implicitly put back into the internal
energy.
We also explicitly include the non-adiabatic effect due to
the field-aligned thermal conduction (eqs. [\ref{eqetot}] and [\ref{conductflux}])
in the energy equation, but in a limited way.
Given the temperature and density in the solar corona, the field aligned thermal
conduction would pose a severely limit on the time step of numerical
integration.  Here we artificially limit the conductive heat flux by limiting
$(\gamma -1) (m_p / 2 k \rho) \kappa_0 T^{5/2}$ to a maximum value of
$\delta x^2 / (8 \, \delta t)$, where $m_p$ is the proton mass, $k$ is the
Boltzman constant, $\delta x$ is the smallest dimension of the grid cell,
and $\delta t$ is the numerical time step calculated from the Courant condition
based on all the other dynamical terms. Note, this limiting is a purely
numerical consideration to avoid the extremely stringent time step that would
be required by the Courant condition set by the thermal diffusion at large
$r$ where density decreases rapidly.  This limiting
is mainly altering thermal conduction in the region above
$r=1.4 R_{\odot}$, to reduce it such that the time step set by the Courant
condition for the thermal diffusion does not go below that set by
the fastest dynamical speed.
Since we are focusing on the thermal
signatures that develop during the quasi-static phase over a time scale long
compared to the dynamic time scale and inside the flux rope below
$r= 1.4 R_{\odot}$ where the thermal conduction is unaltered, we do not
expect this limiting of thermal conduction to significantly alter the
qualitative properties of the thermal signatures.
Thus with the present treatment of the thermodynamics,
we aim to {\it qualitatively} identify regions of significant heating
due to the formation and dissipation of current layers and the
resulting regions of enhanced temperature with the redistribution of heat
by the field aligned thermal conduction.

We solve the above compressible MHD equations using a numerical code which
has its origin from the ZEUS3D code of Stone and Norman (1992a,b), but has
been substantially modified and rewritten, in particular in the schemes used
to solve the continuity, momentum and the energy equations, and in the scheme
for time stepping \citep{fan2009, fan2011}.
For ease of future reference, we call this code ``MFE'' standing for
``Magnetic Flux Eruption'', since it has been mainly used for modeling flux
emergence from the interior into the solar atmosphere and corona in Cartesian
geometry \citep{fan2009}, and for modeling CME initiation in the corona in
spherical geometry \citep{cottaar_fan2009,fan2010,fan2011}.
Here we summarize the numerical schemes used by the MFE code for solving the
above MHD equations.  The dependent field variables are discretized
using a staggered mesh in the spherical domain with $r$, $\theta$, $\phi$
coordinates \citep{stone_norman1992a}.  A modified, second order accurate
Lax-Friedrichs scheme described in \citet[][see eq. (A3) in that paper]
{rempeletal2009} is applied for
evaluating the fluxes in the continuity and the energy equations.
The standard second order Lax-Friedrichs scheme is used for evaluating
the fluxes in the momentum equation.  A method of characteristics that is
upwind in the Alfv\'en waves \citep{stone_norman1992b} is used for evaluating
the ${\bf v} \times {\bf B}$ term in the induction equation, together with
the constrained transport scheme to ensure $\nabla \cdot {\bf B} = 0 $ to
the machine precision.
For temporal discretization, the code uses a second-order accurate
predictor-corrector time stepping.
The thermal conduction term is treated
via an operator splitting procedure as follows. All the other equations and the
total energy equation (eq. [\ref{eqetot}]) excluding the thermal conduction
term are first advanced one full time step (with the second-order
predictor-corrector time stepping). Then a separate
step (with a second-order predictor-corrector time stepping) is carried out
to further update the internal energy for the thermal conduction term and
adjust the total energy accordingly.

\begin{figure}
\epsscale{1.}
\plotone{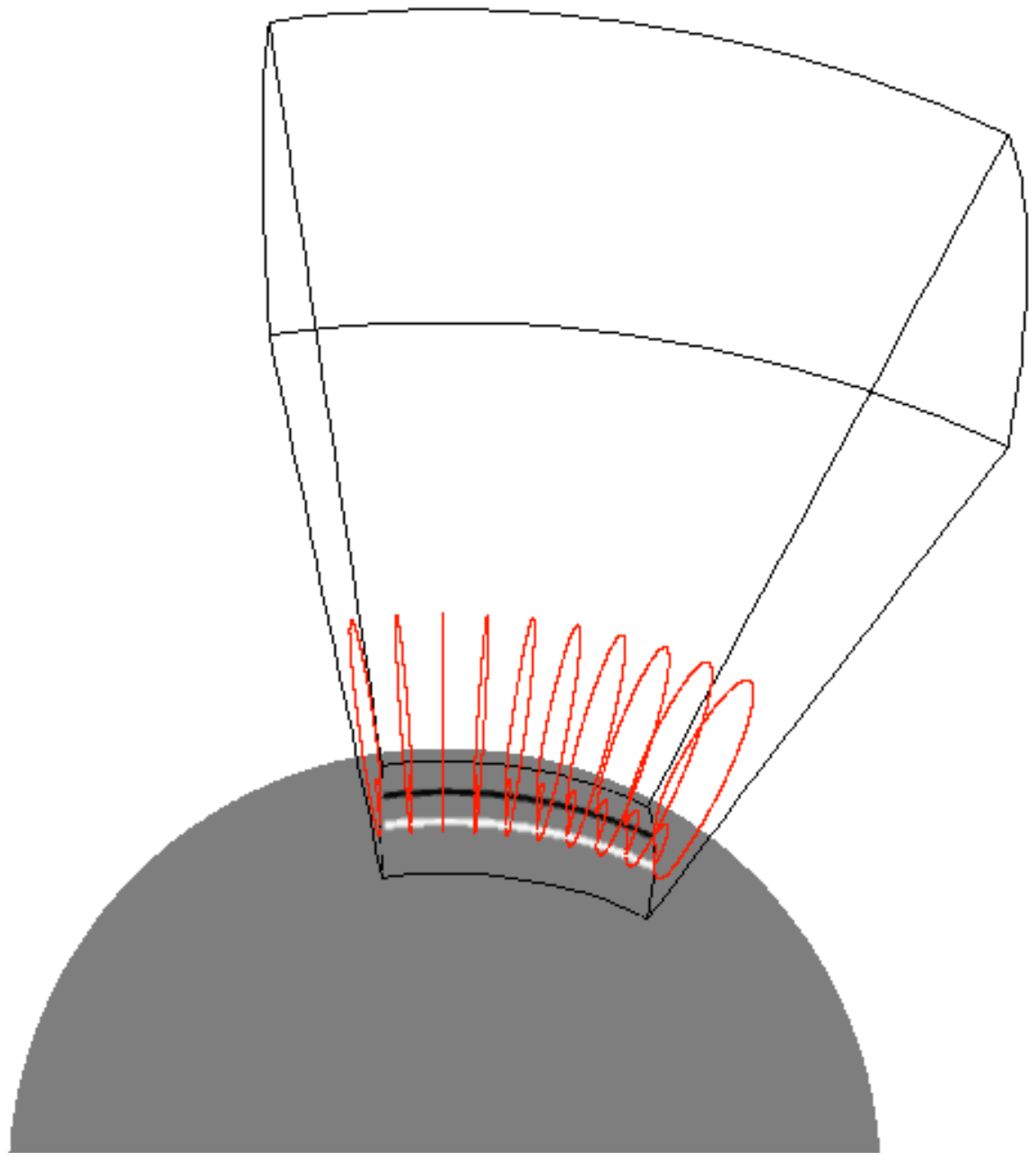}
\caption{The initial configuration of the simulation, same as that used in
\citet{fan2010}, see text for details.} 
\label{fig_init}
\end{figure}

The initial set up of the simulation is the same as that of \citet{fan2010}.
The spherical domain (see Figure \ref{fig_init}) representing the corona is
given by $r \in [R_{\odot}, 5.496 R_{\odot}]$,
$\theta \in [ 5 \pi /12, 7 \pi / 12]$, $\phi \in [- \pi /9.6, \pi /9.6]$,
where $R_{\odot}$ is the solar radius.
It is resolved by a grid of $432 \times 192 \times 240$, which is
uniform in $\theta$ and $\phi$, and non-uniform in $r$: in the range from
$r= R_{\odot}$ to $r=1.788 R_{\odot}$, the grid size is
$dr = 0.0027271 \, R_{\odot} = 1.898 \, {\rm Mm}$,
and $dr$ increases gradually for
$r>1.788 \, R_{\odot}$, reaching about $dr=0.09316 \, R_{\odot}$ at the
outer boundary.

Initially, the domain is assumed to be in hydrostatic
equilibrium at a uniform temperature of $T_0 = 1$ MK with the density and pressure
given by:
\begin{equation}
\rho = \rho_0 \exp \left( - \frac{R_{\odot}}{H_{p0}}
\left ( 1 - \frac{R_{\odot}}{r} \right ) \right )
\end{equation}
\begin{equation}
p = \frac{R T_0 \rho}{\mu},
\end{equation}
where $\rho_0 = 8.365 \times 10^{-16}$ g cm$^{-3}$ is the initial density at
the coronal base, and $H_{p0} = (R T_0 / \mu) ( G M_{\odot}/R_{\odot}^2 )^{-1}$
denotes the initial pressure scale height, which is about $60$ Mm.
The initial atmosphere contains a pre-existing potential arcade
field, whose normal field
$B_r(0,\theta)$ at the lower boundary (see the gray scale image on
the sphere in Figure \ref{fig_init}) is given by
\begin{equation}
B_r(0, \theta) = \frac{1}{R_{\odot}^2 \sin \theta}
\frac{d A_s (\theta)}{d \theta},
\end{equation}
where,
\begin{equation}
A_s (\theta) =
\left \{ \begin{array}{ll}
0,
&
\frac{5}{12} \pi < \theta < \frac{\pi}{2} - \theta_t - \theta_a, \\
- \frac{\theta_a}{\pi} \sin \theta_p
B_0 R_{\odot}^2
\left[ 1 - \cos \left[ \frac{\pi}{\theta_a}
\left[ \theta - \left( \frac{\pi}{2} - \theta_a - \theta_t \right) \right]
\right] \right],
&
\frac{\pi}{2} - \theta_t - \theta_a < \theta <
\frac{\pi}{2} - \theta_t, \\
- \frac{2 \theta_a}{\pi} \sin \theta_p
B_0 R_{\odot}^2,
&
\frac{\pi}{2} - \theta_t < \theta < \frac{\pi}{2} + \theta_t, \\
- \frac{\theta_a}{\pi} \sin \theta_p
B_0 R_{\odot}^2
\left[ 1 + \cos \left[ \frac{\pi}{\theta_a}
\left[ \theta - \left( \frac{\pi}{2} + \theta_t \right) \right ]
\right ] \right ],
&
\frac{\pi}{2} + \theta_t < \theta < \frac{\pi}{2} + \theta_t + \theta_a, \\
0,
&
\frac{\pi}{2} + \theta_t + \theta_a < \theta < \frac{7}{12} \pi ,
\end{array} \right .
\end{equation}
in which $\theta_a = 0.05$, $\theta_t = 0.0432$, $\theta_p = \pi / 2 -
\theta_t - \theta_a / 2$, and $B_0 = 20 {\rm G}$ is the peak field
strength in the arcade field.
Thus the peak Alfv\'{e}n speed at the foot point of the arcade field
is $v_{A0} = B_0 / \sqrt{4 \pi \rho_0} = 1951 $ km ${\rm s}^{-1}$
which is more than a factor of 10 greater than the initial sound
speed of $135$ km ${\rm s}^{-1}$.

As was in \citet{fan2010}, we impose (kinematically) at the lower boundary
of the domain (at $r=R_{\odot}$) the emergence of a twisted magnetic torus
${\bf B}_{\rm tube}$, by specifying a time dependent transverse electric field
${\bf E}_{\perp}|_{r=R_{\odot}}$ that corresponds to the upward advection
of the flux tube at a velocity ${\bf v}_0$:
\begin{equation}
{\bf E}_{\perp}|_{r=R_{\odot}} = {\hat{\bf r}} \times \left [ \left (
- \frac{1}{c} \, {\bf v}_0 \times
{\bf B}_{\rm tube} (R_{\odot}, \theta, \phi, t) \right )
\times {\hat{\bf r}} \right ].
\label{eq_emf}
\end{equation}
Here the imposed velocity field on the lower boundary is a constant
${\bf v}_0$ in the area where the emerging tube intersects the lower
boundary and zero in the rest of the area.
The field structure ${\bf B}_{\rm tube}$ used for specifying
${\bf E}_{\perp}|_{r=R_{\odot}}$ is an axisymmetric torus
defined in its own local spherical polar coordinate system
($r'$, $\theta'$, $\phi'$) whose origin is located at
${\bf r} = {\bf r}_0 = (r_0, \theta_0, \phi_0)$ of the sun's spherical
coordinate system and whose polar axis is
parallel to the polar axis of the sun's spherical coordinate system:
\begin{equation}
{\bf B}_{\rm tube} = \nabla \times \left (
\frac{A(r',\theta')}{r' \sin \theta' } \hat{\bf \phi'} \right )
+ B_{\phi'} (r', \theta') \hat{\bf \phi'},
\end{equation}
where
\begin{equation}
A(r',\theta') = \frac{1}{2} q a^2 B_t
\exp \left( - \frac{\varpi^2(r',\theta')}{a^2} \right) ,
\label{eqafunc}
\end{equation}
\begin{equation}
B_{\phi'} (r', \theta') = \frac{a B_t}{r' \sin \theta'}
\exp \left( - \frac{\varpi^2(r',\theta')}{a^2} \right),
\label{eqbph}
\end{equation}
where $a$ is the minor radius of the torus,
$\varpi = (r'^2 + R'^2 -2r'R' \sin \theta')^{1/2}$ is
the distance to the curved axis of the torus, with $R'$ being the major
radius of the torus, $q$ denotes the angle (in rad) of field
line rotation about the axis over a distance $a$ along the axis, $B_t a/R'$
gives the field strength at the curved axis of the torus.
Here we have $a = 0.04314 R_{\odot}$, $R' = 0.25 R_{\odot}$,
$q/a = - 0.0166$ rad ${\rm Mm}^{-1}$, $B_t a/R = 2.24 B_0$.
The magnetic field ${\bf B}_{\rm tube}$ is truncated to zero outside of the
flux surface whose distance to the torus axis is $\varpi = a$.
For specifying the flux emergence via ${\bf E}_{\perp}|_{r=R_{\odot}}$
given by (\ref{eq_emf}), it is assumed that the torus' center is initially
located at ${\bf r}_0 = (r_0 = 0.707 R_{\odot}, \, \theta_0 = \pi /2, \,
\phi_0 = 0) $ (thus the torus is initially entirely below the surface),
and it moves bodily towards the lower boundary at a constant
velocity ${\bf v}_0 = v_0 \hat{{\bf r}}_0$, with $v_0 = 0.001 v_{A0}$,
until a time $t_{\rm stp}$, when the emergence is stopped and
${\bf E}_{\perp}|_{r=R_{\odot}}$ is set to zero.  During flux emergence,
we assume that the density inside the emerging flux tube is uniformly
$\rho_0$ (same as the initial density at the bottom of the domain).
Thus as the tube is being transported into the domain via
the electric field at the lower boundary, there is also an inflow
of mass flux $\rho_0 v_{0r}$ through the lower boundary in the
area where the emerging tube intersects the boundary.
When the emergence is stopped, both the electric field and the velocity at
the lower boundary are set to zero, with no inflows or
outflows and with field line foot points rigidly anchored.
We assume perfectly conducting walls for the side boundaries of the simulation
domain.  For the outer boundary, we use a simple outward extrapolating
boundary condition that allows plasma and magnetic field to flow through.
These boundary conditions described above are the same as those used in
\citet{fan2010}

For the present simulation, we drive the emergence of the torus via the
electric field at the lower boundary until
$t_{\rm stp} = 90 R_{\odot} / v_{A0}$,
the same as case e4 in \citet{fan2010}. However the difference between this
simulation and those in \citet{fan2010} is that here we no longer assume
the simple isothermal evolution, but instead we solve the total energy
equation for a perfect gas with an adiabatic index $\gamma = 1.1$ for
the coronal plasma. Thus we take into
account non-adiabatic effects due to heating produced by (numerical)
dissipation of kinetic and magnetic energies. We also incorporate
thermal conduction along the magnetic field lines, as is described earlier
in \S \ref{sec:model}.
Thus temperature can vary due to adiabatic expansion/compression, as well as
due to the non-adiabatic effects: heating and thermal conduction.
This enables us to identify sites of significant heating due to
current sheet formation and study the resultant temperature distribution
(qualitatively) during the evolution of the 3D flux rope in the corona.

\section{Results\label{sec:result}}
\begin{figure}
\epsscale{0.275}
\plotone{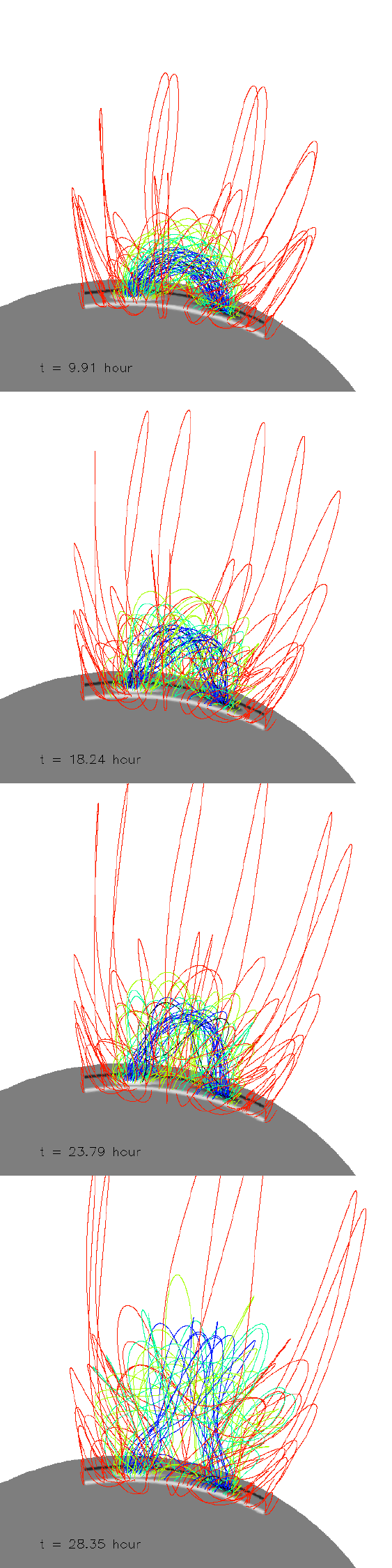}
\caption{Snapshots of the 3D coronal magnetic field lines after the imposed
flux emergence is stopped.  A movie of the magnetic field evolution together
with the evolution shown in Figure \ref{fig_merievol} below is also
available in the electronic version of the paper}
\label{fig_3devol}
\end{figure}

\begin{figure}
\epsscale{0.75}
\plotone{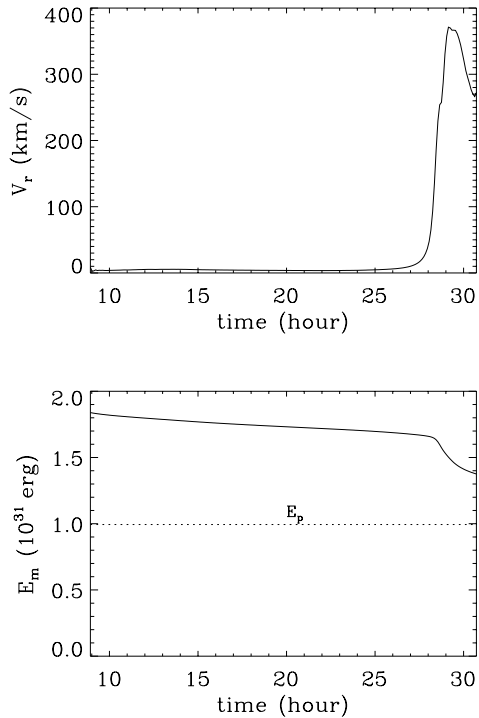}
\caption{Rise velocity at the apex of the tracked axis of the emerged flux
rope (upper panel) and the total magnetic energy $E_m$ in the simulation domain
(lower panel) as a function of time after the imposed flux emergence has
stopped. The dotted line marks the magnetic energy of the corresponding
potential field, $E_p$, which remains unchaged after the emergence has
stopped. The eruption causes a roughly 43\% reduction of the free magnetic
energy.}
\label{fig_vr_em}
\end{figure}

Figure \ref{fig_3devol} shows snapshots of the 3D evolution of the coronal
magnetic field after the imposed flux emergence at the lower boundary has
stopped at $t_{\rm stp} = 90 R_{\odot} / v_{A0} = 8.92 $ hour.
Figure \ref{fig_vr_em} shows the temporal evolution of the rise speed
at the apex of the tracked axis of the emerged flux rope (upper panel) and
the total magnetic energy in the coronal domain (lower panel) after
$t_{\rm stp}$. The way we track the axis of the emerged flux rope is the same
as that described in \citet[][see discussion under Figure 3 in that
paper]{fan2010}.  We find that the flux rope first settles into a
phase of quasi-static rise, and then begins to accelerate and erupt when
a critical height of roughly
$1.25 R_{\odot}$ is reached by the tube axis.
During the quasi-static phase after the emergence is stopped,
there is no more input of Poynting flux from the lower boundary and the
magnetic energy in the domain is slowly
decreasing (see Figure \ref{fig_vr_em}) due to tether cutting reconnections
in the current sheets that develop, as we will discussed later.
Compared to the corresponding
case e4 in \citet{fan2010} with the same imposed lower boundary conditions,
the present case with a different treatment of the thermodynamics
is found to undergo a longer quasi-static rise phase before the flux rope
reaches roughly the same critical height for the onset of the torus instability.
As a result, at the time of the onset of eruption, the (free) magnetic energy
is lower and the Alfv\'en speed at the rope axis
is also lower (by about 15\%) in the present case. Thus it is found that
the eruption is slightly less energetic with the flux rope accelerating
to a lower peak speed of about $371 {\rm km/s}$ for the eruption, compared
to the peak speed of about $429 {\rm km/s}$ in the corresponding e4 case in
\citet{fan2010}.
Here we focus on studying the development of the thermal
signatures in the flux rope as a result of current sheet formation and
tether-cutting reconnections during the quasi-static rise phase.
The tether-cutting reconnections after the emergence is stopped
continue to build up the twisted flux of the flux rope (as computed
in equations [14] in \citet{fan2010}), even no more Poynting flux is
coming through the lower boundary and the magnetic energy is declining.
They enable the flux rope to reach the
critical height and eventually erupt. Therefore the thermal structures
that develop due to the reconnections may be important signatures 
for indicating the readiness of the flux rope to erupt, and identifying
such thermal signatures in the observables is useful.

Figure \ref{fig_merievol} shows snapshots of the density
(in log scale), current density normalized by magnetic
field strength ($J/B \equiv | \nabla \times {\bf B} | / B$),
and temperature in the central meridional
cross-section of the flux rope shown in the corresponding snapshots
in Figure \ref{fig_3devol}.
In the density cross-section during the quasi-static rise phase
(panels in the left column), the outermost low density layer
corresponds to the pre-existing arcade fields, and the low
density is produced by the stretching and expansion of the pre-existing arcade
fields immediately surrounding the emerged flux rope fields.
Inside the outermost low density layer is a layer of relatively high density
and high $J/B$ (indicated by the yellow arrows in Figure \ref{fig_merievol}),
which corresponds to the outer boundary of the emerged flux rope.
Inside the flux rope, there is a central low density void (indicated
by the green arrows in Figure \ref{fig_merievol}) which grows in size during
the quasi-static phase.  The central void is closed at the bottom with a
sharp ``U'' shaped lower
boundary, and is enclosed in a high density shell.  The sharp
lower boundary of the void corresponds to a ``U'' shaped current concentration 
of high $J/B$ which extends from a strong central vertical current layer below
(see the middle and bottom snapshots in the 2nd column of
Figure \ref{fig_merievol}).
In other words, the central void is shaped like a ``lollipop'' on top of a
vertical ``stick" corresponding to the central vertical current layer, and
the current layer extends upward to surround the
two sides of the void.
The middle and lower images in the right column of Figure
\ref{fig_merievol} show
that the central growing void is the hottest part in the central meridional
cross-section, hotter than the plasma in the strongest vertical current layer
below it and the current layer surrounds it.
Furthermore, this hot void is growing below the apex of the
emerged flux rope axis marked with the yellow ``x'' point in the
cross-sections shown in Figure \ref{fig_merievol}.

Figure \ref{fig_3dfdlcurr} shows the 3D coronal field lines and the isosurface
of $J/B$ at a value of $1/(10 \, dr)$ where $dr$ is the smallest grid size in
$r$.  We find that at $t=9.91 $ hour, corresponding to the top row of Figure
\ref{fig_merievol}, the strongest portions of the current layers are two ``J''
shaped layers curving about the two legs of the flux rope at the boundaries
between the flux rope legs and the arcade with the opposite polarity
foot points.
This is why the central meridional cross-section of $J/B$ (the top panel in the
middle column of Figure \ref{fig_merievol}) shows a double legged
structure, with each leg belonging to one of the ``J''-layers.
When the ``lollipop'' shaped central hot void
has developed in the central meridional cross-section (as is shown
in the middle and bottom rows of Figure \ref{fig_merievol} for
$t=18.24$ hour and $t=23.79$ hour), the strongest portion of the current
layer has become a central
inverse S-shaped vertical layer underlying the flux rope (see the middle
and right panels of Figure \ref{fig_3dfdlcurr}, and the associated movie
in the electronic version).
The inverse-S shape is consistent with the left-handed twist of the flux
rope \citep[e.g.][]{fan_gibson2004}.
A forward-S shape for the current layer would result for
a right-hand twisted flux rope.

We find that these current layers in the simulation (as shown in
Figure \ref{fig_3dfdlcurr}) develop along topological structures identified
as Quasi Separatrix Layers (QSLs), which are regions of the magnetic volume
where the field line connectivity experiences drastic
variations \citep{demoulinetal1996a, demoulinetal1996b, titovetal2002}.
They are a generalization of the concept of
separatrices at which the field line linkage is discontinuous.
Similar to a separatrix, a QSL divides the coronal
domain into quasi-connectivity domains, and due to the drastic change of
the field line connectivity, it is a site along which thin and intense
current sheets tend to build up because of the distinct dynamic
behaviors and evolutions of the
field lines in each of these subdomains in the line-tied corona
\citep[e.g.][]{demoulinetal1996b,aulanieretal2005,savchevaetal2012b}.
QSLs are located by estimating the so-called squashing degree, $Q$, which
is defined as the square of the norm of the Jacobian matrix of the field
line mapping from one foot point to the other, divided by the absolute value
of the determinant of the Jacobian matrix \citep{titovetal2002,titov2007,
pariat_demoulin2012}. QSLs correspond to regions of very large
$Q$, larger than the averaged values of the rest of the domain by many orders
of magnitude \citep[e.g.][]{pariat_demoulin2012,
savchevaetal2012b}.  We have evaluated the squashing
degree $Q$ in our simulated coronal magnetic field using the
formulation described in \citet[][method 3 in the paper]{pariat_demoulin2012}.
The left panels in Figure \ref{fig_qfac} show the computed $Q$ in respectively
the central meridional cross-section (upper panel) and the cross-section at
the height of $r=1.048 R_{\odot}$ (lower panel) in the 3D magnetic field shown
in the right panel of Figure
\ref{fig_3dfdlcurr}).  Compared to the corresponding 2D cuts of $J/B$ shown
in the right panels of Figure \ref{fig_qfac}, we can see that the thin, most
intense current layers indeed form along the QSLs with the largest $Q$ values,
which are orders of magnitude greater than the averaged $Q$ in the rest of the
domain. The QSLs with the highest $Q$ values shown in the central
meridional cross-section (top left panel of Figure \ref{fig_qfac})
correspond to the
mid cross-section of the so-called Hyperbolic Flux Tube (HFT), a
generalization of the X-line configuration, which divides the magnetic
volume into 4 distinct domains of magnetic field connectivity
\citep[e.g.][]{titov2007,aulanieretal2005,savchevaetal2012b}.
The main central vertical
current layer shown in the right panel of Figure \ref{fig_3dfdlcurr},
(whose cross-sections are shown in the 2D cuts in Figure \ref{fig_qfac}),
forms along the HFT and is likely a
thin current sheet that can lead to significant reconnection even under the
realistic high Lundquist number condition of the solar corona.

To understand the 3-dimensional structure of the growing central hot void
on top of the central vertical current layer seen in the
cross-sections in Figure \ref{fig_merievol}, we show in the top panels of
Figure \ref{fig_hotchannel_t2} the horizontal cross-sections of density and
temperature at $r=1.15 R_{\odot}$ for $t=23.79$ hour (corresponding to the
height indicated by the green arrows in the bottom row of
Figure \ref{fig_merievol}).  We see a hot channel of inverse-S shape, with the
main middle segment of the channel tilted away from the east-west
direction (or the direction of the emerging flux rope axis) clock-wise by
roughly $40^{\circ}$.  There is a good anti-correlation between the
temperature and the density of the channel, i.e. it is a hot, low-density
channel. In the lower panels of Figure \ref{fig_hotchannel_t2}, we also show
the 3D morphology of the hot channel as outlined by
the pink isosurface of temperature (at 1.2 MK) in relation to the current
layers (orange isosurfaces of $J/B$) and the magnetic field lines.
The hot channel is on top of the central vertical current layer, and the
current layer extends upward to surround the two sides of the hot channel
in the middle portion (see the movie associated with
Figure \ref{fig_hotchannel_t2} in the electronic version which show the
rotating view of the 3D structures). The hot channel also
threads under the axial field line of the flux rope (as indicate in
the bottom panels of Figure \ref{fig_merievol}).

If this flux rope is viewed above the
limb, with a line of sight that is aligned with the main middle segment
of the hot channel to a suitable angle, then one would be able to observe
the central void of high temperature inside the flux rope.
Figure \ref{fig_limbview_40} shows the modeled images above the limb of
the emission measure (panel b), line-of-sight averaged temperature (panel c),
and SDO/AIA intensities for the $171$ {\AA} (panel d), $193$ {\AA} (panel e),
and $211$ {\AA} (f) filters, by integrating through the
simulation domain along the line-of-sight that is tilted $40^{\circ}$
clock-wise from the east-west direction (or the direction of the axis of the
flux rope).  Also shown in the figure is the view above the limb of the
3D field lines and the iso-surface of $J/B$ (at a value of
$1/10 \, dr$ with $dr$ being the smallest radial grid size) along the same
line of sight (Figure \ref{fig_limbview_40}a).
For computing the modeled emission above the limb
at the AIA $171$ {\AA}, $193$ {\AA}, and $211$ {\AA} wavelength channels,
we use the temperature and density from the simulation
and carry out the following integration along the line-of-sight $l$ through
the domain:
\begin{equation}
I_i = \int n_e^2 (l) \, f_i (T(l), n_e(l)) \, dl,
\end{equation}
where the subscript $i$ denotes one of the AIA wavelength channels, 
$I_i$ denotes the intensity at each pixel in units of DN/s/pixel,
$n_e$ is the electron number density,
and the function $f_i (T,n_e)$ takes into account the atomic physics
and the properties of the AIA filters. We have ignored the weak
dependence of $f_i$ on $n_e$, and obtained the temperature dependent
function $f_i (T)$ using the SolarSoft routine {\tt get\_aia\_response.pro}.

The resulting modeled emission measure and the AIA intensity images in
Figure \ref{fig_limbview_40} show
a central tear shaped void with a relatively sharp ``U'' shaped
lower boundary,
surrounded by a relatively dense shell. The central tear shaped void is
the central low density, hot channel viewed along its main part of the
length (see Figure \ref{fig_hotchannel_t2}), and the ``U'' shaped dense
shell enclosing the void is produced by the upward
extensions of the vertical current layer (see the orange
isosurfaces in Figures \ref{fig_limbview_40}a) which has formed along
the HFT.  Note that the tear-shaped void enclosed by a ``U'' shaped dense
shell on top of a vertical dense structure is inside the larger flux rope
structure.
Such a central density sub-structure inside the flux rope
has been observed in some of the quiescent prominence cavities at the limb
\citep[e.g.][]{regnieretal2011, reevesetal2012, berger2012}.
Figure \ref{fig_obsexample} shows a prominence cavity observed by SDO/AIA
on 12 June 2012 at about 23:01 UT on the north-west limb.
This is also the case studied by \citet[][see also Figure 1 in that paper
for the sharper negative images] {regnieretal2011}.
The AIA 171, 193, 211 images all show an ellipse shaped central void with
a sharp ``U'' shaped lower boundary situated on top of the vertical
prominence.
The void is enclosed by a ``U'' shaped dense shell extending upward
from the prominence. These qualitative features are similar to those
seen in the modeled AIA images shown in Figure \ref{fig_limbview_40},
provided that the observed prominence material is co-spatial with
the vertical current layer in the model
(Figure \ref{fig_limbview_40}a).
Prominence condensation may develop inside the denser current layer due to
radiative cooling, which is not included in our numerical simulation.
The ``U'' shaped dense shell extending upward from the prominence seen
most prominently in the AIA 171 observations is also called ``horns''
\citep{berger2012} and is found to be common features of quiescent
prominences observed with the AIA.
Our model suggests that this observed density feature of a dense
prominence column with upward extending ``horns''
corresponds to the current sheet that forms along
the HFT, when viewed from a suitable line of sight angle, as illustrated
by Figure \ref{fig_horns}, which can be compared qualitatively with the
the prominence and ``horn'' morphology shown in Figure 3 of \citet{berger2012}.
In other words, the development of a dense prominence column with
upward extending ``horns'' enclosing a
central void (such as the cases shown in \citet{regnieretal2011} and
\citet{berger2012}) signifies the development of the HFT topology
(and the associated tether cutting reconnections) in the cavity flux rope.
Figure \ref{fig_horns} also shows two field lines drawn from two points in
close proximity on opposite sides of one of the upward extensions of
the central current layer. It illustrates the drastic difference in the
connectivities of these two field lines which come to close proximity
on opposite sides of the current layer extension
(which is a QSL) with very different field line orientations.
A 3D rotating view of the current layer and
the field lines is available as a movie associated with Figure \ref{fig_horns}
in the online version.

The anti-correlation between the emission measure and the temperature for
the central void structure (Figures \ref{fig_limbview_40}(b)(c)
) is also consistent with the recent study
of internal thermal properties of coronal cavities by \citet{reevesetal2012}.
They found a central core with lower density but higher temperature inside the
larger coronal cavity.  This central hot core is also consistent with the
X-ray emitting core sometimes observed in coronal cavities \citep[e.g.][]
{hudsonetal1999,reevesetal2012}.

Our model suggests that this sub-structure of the hot, low
density channel inside
the larger flux rope corresponds to the reconnected, twisted flux being added
to the flux rope via tether-cutting reconnections in the current layers
which have formed along the HFT during the quasi-static stage
(see orange surfaces in Figure \ref{fig_3dfdlcurr}).
Figure \ref{fig_reconnect}(b) illustrates such a reconnected field line
through the central hot void (see the red field line), which at $t=23.79$ hour
has become a field line that twists about the tracked axis (black field line)
of the flux rope. Yet, this red field line has its foot points rooted in the
arcade regions.  At an earlier time $t=9.91$ hour (see
Figure \ref{fig_reconnect}(a)), the field lines drawn from
these foot points are two simple arcade field lines (the red field lines
in Figure \ref{fig_reconnect}(a)).  Thus the two simple arcade field lines
at $t=9.91$ hour have undergone multiple reconnections with the flux rope
fields, and by time $t=23.79$ hour have turn into a field line rooted in the
arcade foot points but twists about the axis of the flux rope. Such addition
of the twisted flux to the flux rope, threading under the axis, produces
the expanding, low density hot channel in the flux rope.
When viewed in a line of sight that is roughly
aligned with the channel, the resulting central void
is found to enlarge and rise up during the
quasi-static stage (see Figure \ref{fig_limbview_evol} and the associated
movie in the online version of the paper).  As the critical height is reached,
the central void accelerates rapidly upward and is ejected, leaving behind
a cusp shaped brightening (see the last row in Figure \ref{fig_limbview_evol}).
The development of the long bright current sheet trailing the ejected central
void (see the movie), and the rise of the cusp shaped brightening
are more prominent in the 193 and 211 channels due to their sensitivity to
higher temperatures.
This evolution sequence can be compared with the observed evolution of
the prominence cavity studied in \citet{regnieretal2011}, which begins to
erupt at about 03:24 UT on June 13 2010.  A composite movie of the AIA
observation in the 171, 193 and 211 channels of the evolution of this
cavity on the
north-west limb during the period from 18:00 UT 12 June 2010 to 21:30 UT 13
June 2010 can be viewed at
\url{http://sdowww.lmsal.com/suntoday/index.html?suntoday_date=2010-06-12#}.
The observed evolution show many qualitative similarities with the modeled
one, including the morphology of the central void enclosed in the U-shaped
dense shell on top of a dense column during the quasi-static phase, and
the ejection of the void and the post-eruption brightening.

\section{Conclusions\label{sec:conc}}
We have carried out an MHD simulation of the quasi-static rise and the onset
of eruption of an anchored coronal flux rope.  Earlier in the simulation, the emergence of a twisted flux rope is imposed at the lower boundary into a
pre-existing coronal potential arcade field. Then the emergence is stopped at
the lower boundary and the coronal magnetic field is allowed to settle into
a quasi-static equilibrium.  It is found that the flux rope first settles into
a quasi-static rise phase with an underlying sigmoid-shaped current layer
developing along the HFT, which is argued to be a likely site for the
formation of thin current sheets and magnetic reconnections
\citep[e.g.][]{titov2007, aulanieretal2005, savchevaetal2012b}. Tether cutting
reconnections in the current layer effectively add twisted flux to the flux
rope and allow it to rise quasi-statically (even as the free magnetic energy
is decreasing) and the flux rope undergoes dynamic eruption when it reaches
the critical height for the onset of the torus instability
\citep{aulanieretal2010, fan2010}.
In the present simulation we study the thermal signatures that result from
the development of the HFT topology and the associated tether cutting
reconnections in the the current sheets that form along the HFT.
We solve the total energy equation in conservative form for a perfect gas
with an adiabatic index $\gamma=1.1$, and thus incorporate the non-adiabatic
heating resulting from the dissipation of kinetic and magnetic energies due
to the formation of intense current layers.  Furthermore,
thermal conduction along magnetic field lines is also included in the energy
equation so as to incoporate the redistribution of heat along the field lines.
In this way, we are able to identify qualitatively the thermal
signatures in the flux rope due to heating produced by the tether-cutting
reconnections in the current layers that form during the quasi-static rise
phase of the flux rope 

\begin{figure}
\epsscale{1.}
\plotone{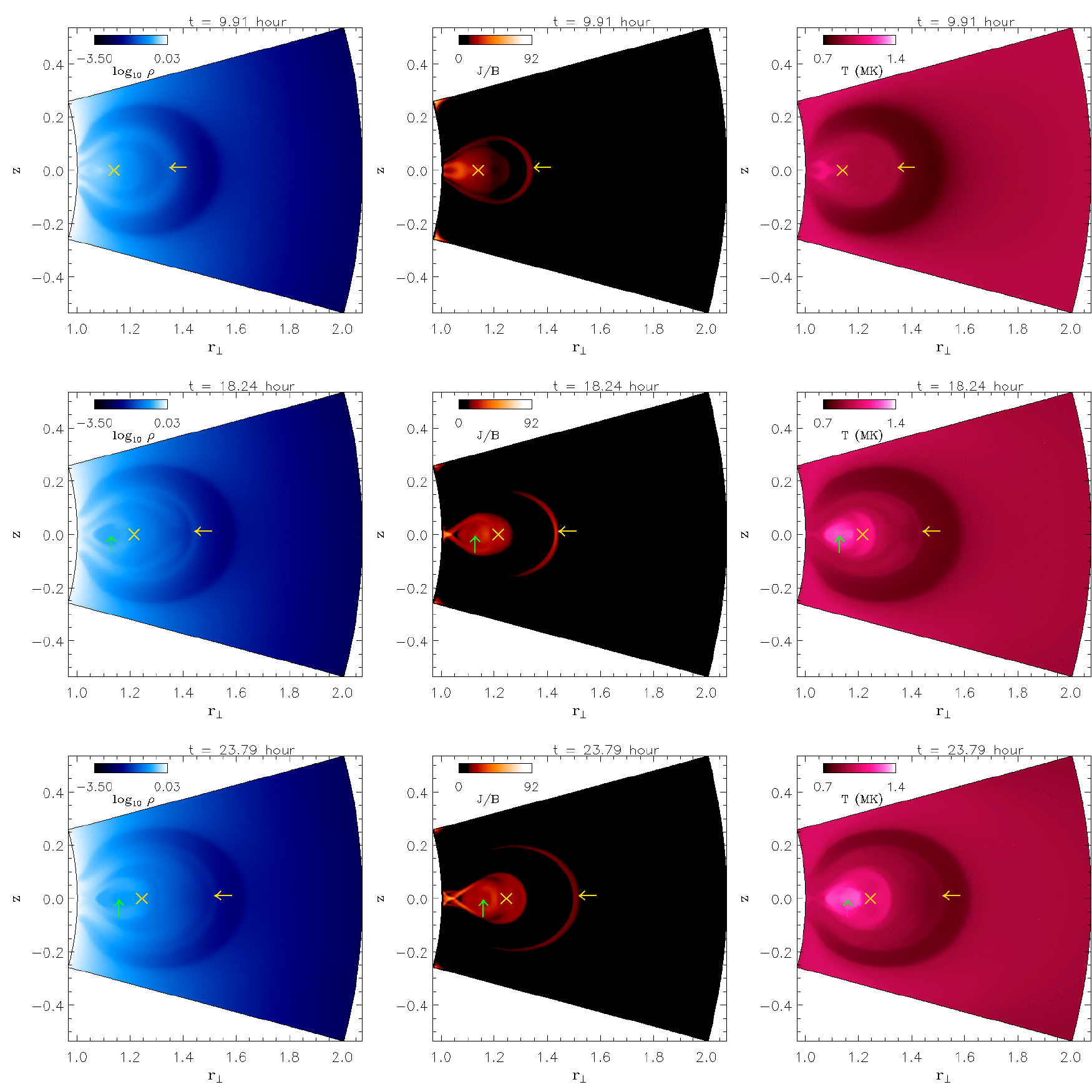}
\caption{Snapshots of the density (in log scale), current
density normalized by magnetic field strength
($J/B \equiv | \nabla \times {\bf B} | / B$), and temperature in
the central meridional cross-section of the flux rope shown in the
corresponding snapshots in Figure \ref{fig_3devol}. Here the density is
in units of $\rho_0 = 8.365 \times 10^{-16}$, and $J/B$ is in units of
$1/R_{\odot}$. The features indicated by the green and the yellow arrows and
the yellow `X' points are discussed in the text.
A movie of the evolution shown in this figure together
with the evolution shown in Figure \ref{fig_3devol} is also
available in the electronic version of the paper}
\label{fig_merievol}
\end{figure}

\begin{figure}
\epsscale{1.}
\plotone{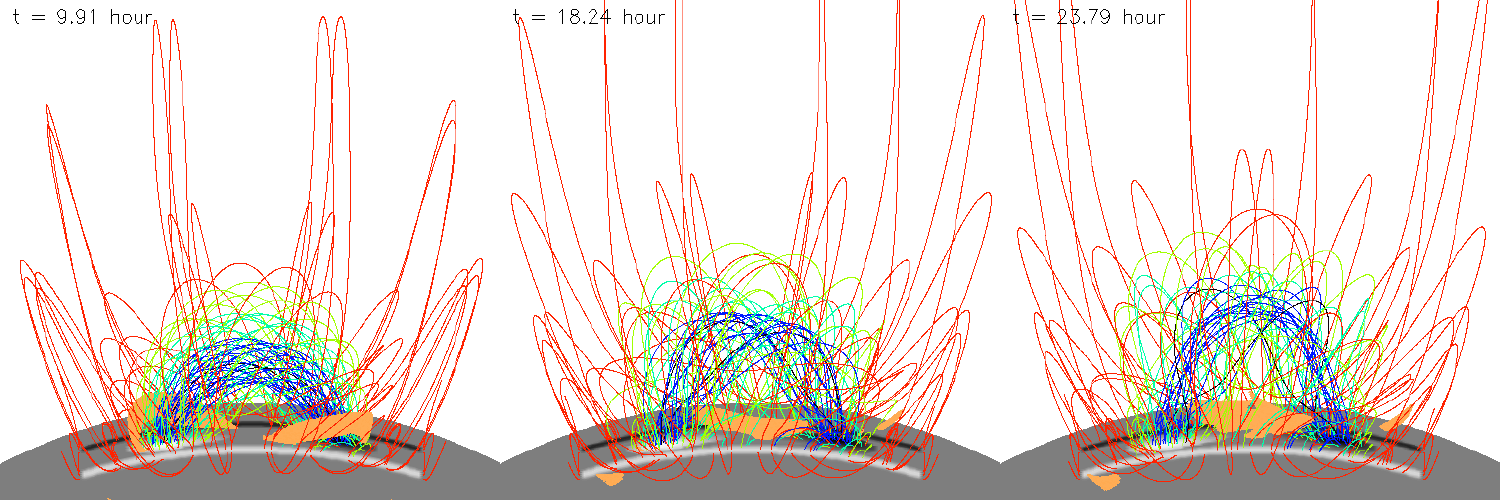}
\caption{The 3D coronal field lines and the iso-surface
of $J/B$ at a value of $1/(10 \, dr)$ where $dr$ is the smallest grid size in
$r$ at times $t=9.91$ hour, $t=18.24$ hour, and $t=23.79$ hour. An animation
of the 3D structures with a rotating perspective is also available in the
electronic version of the paper.}
\label{fig_3dfdlcurr}
\end{figure}

\begin{figure}
\epsscale{1.}
\plotone{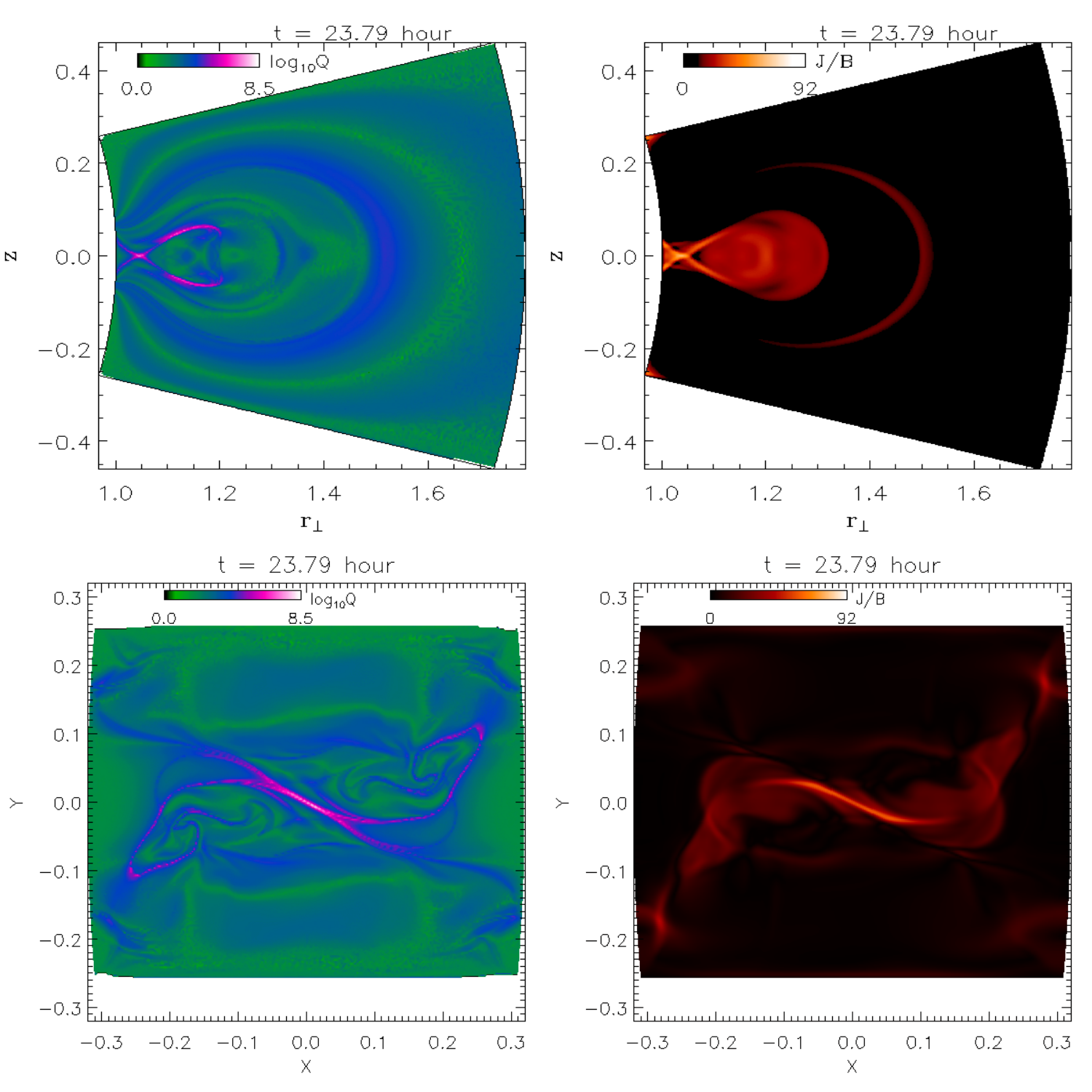}
\caption{Estimation of the squashing degree $Q$ in respectively
the central meridional cross-section (upper left panel) and the
cross-section at the height of $r=1.048 R_{\odot}$ (lower panel) in
the 3D magnetic field shown in the right panel of Figure
\ref{fig_3dfdlcurr}, and the corresponding 2D cuts of $J/B$ (right panels).}
\label{fig_qfac}
\end{figure}

\begin{figure}
\epsscale{1.}
\plotone{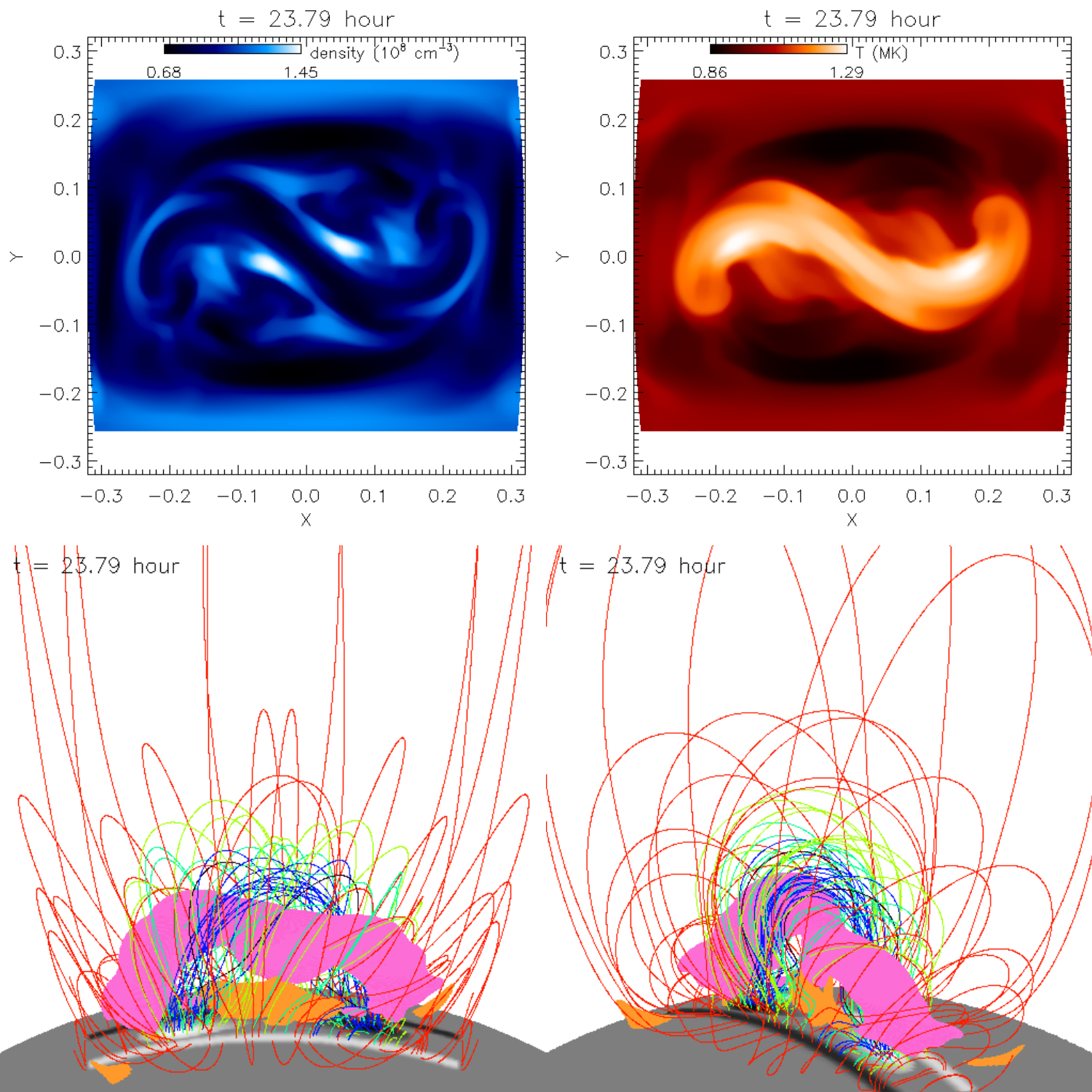}
\caption{Top panels show the horizontal cross-sections of density and
temperature at $r=1.15 R_{\odot}$ for $t=23.79$ hour, corresponding to the
height indicated by the green arrows in the bottom row of
Figure \ref{fig_merievol}. The lower panels show two perspective views of the
3D magnetic field lines, the pink iso-surface of temperuature at 1.2 MK, which
outlines the central hot channel, and the current layers outlined by the
orange iso-surfaces of $J/B$ at the level of $1/(10 dr)$, with $dr$ being
the smallest radial grid size. A movie showing the rotating view of the 3D structures
in the lower panels is available in the electronic version of the paper.}
\label{fig_hotchannel_t2}
\end{figure}

\begin{figure}
\epsscale{1.}
\plotone{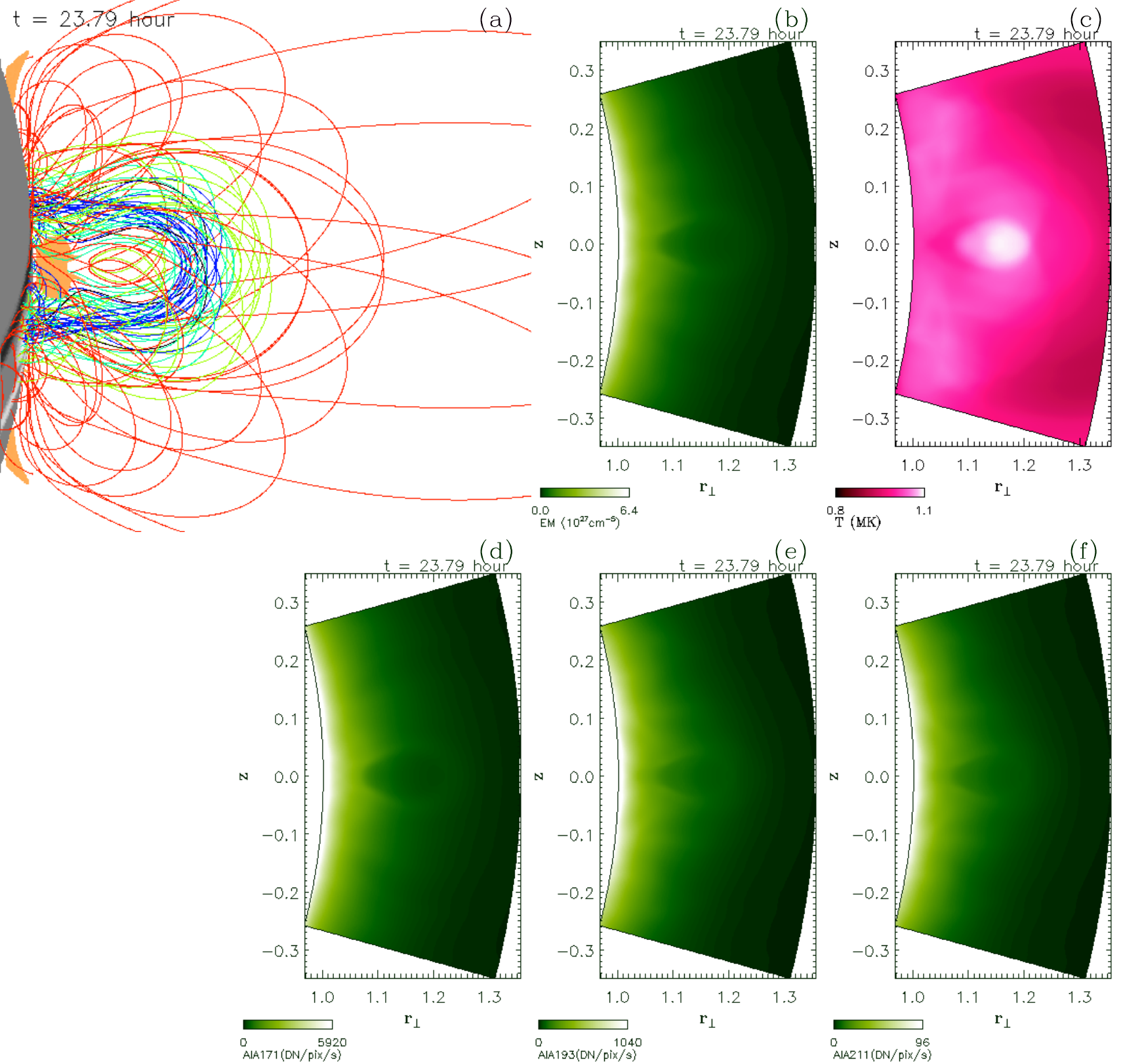}
\caption{(a) The 3D field lines and the iso-surface of $J/B$ at a value of
$1/(10 \, dr)$ with $dr$ being the smallest radial grid size, viewed above
the limb along the line of sight that is tilted by $40^{\circ}$ clock-wise
from the east-west direction (or the direction of the axis of the flux rope).
Also shown are modeled images of the emission measure (b), line-of-sight
averaged temperature (c), and SDO/AIA intensities at $171$ {\AA} (d),
$193$ {\AA} (e), and $211$ {\AA} (f) channels, by integrating through the
simulation domain along the same line-of-sight.}
\label{fig_limbview_40}
\end{figure}

\begin{figure}
\epsscale{1.}
\plotone{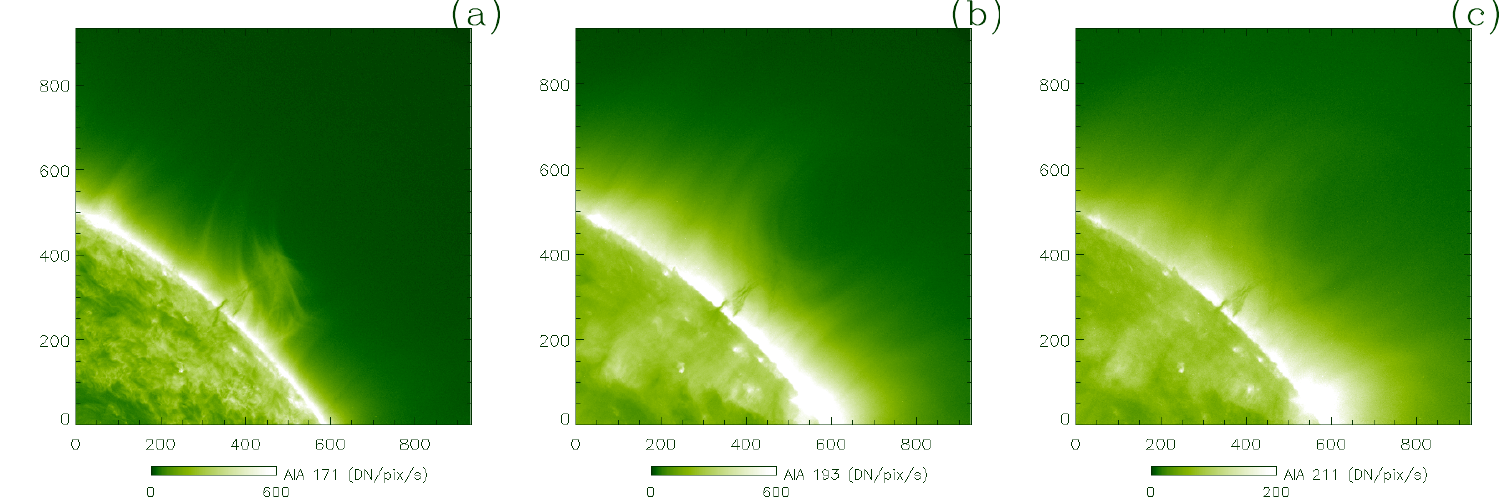}
\caption{A prominence cavity observed by SDO/AIA
on 12 June 2012 at about 23:01 UT on the north-west limb. See also Figure
1 of \citet{regnieretal2011} which shows sharper inverse-color images of the
same cavity at a later time on 13 June 2012 at about 03:36 UT.}
\label{fig_obsexample}
\end{figure}

Intense current layers are found to develop during the quasi-static phase
along QSLs, which are topological boundaries where the magnetic field
connectivity undergoes drastic variations and are likely sites for the
formation of thin current sheets and significant magnetic reconnections
\citep{demoulinetal1996a,demoulinetal1996b, titovetal2002}.
This has been shown also in previous MHD simulations of coronal
flux rope evolution as well as nonlinear force-free field models (NLFFF) of
observed sigmoid active regions
\citep[e.g.][]{aulanieretal2005, aulanieretal2010, savchevaetal2012a,
savchevaetal2012b}.
The strongest portions of the current layers that form are initially two ``J''
shaped layers curving about the two legs of the flux rope at the boundaries
between the flux rope legs and the arcade with the opposite polarity foot points
(see the left panel of Figure \ref{fig_3dfdlcurr}).
Later the strongest portion of the current layer becomes a central
inverse S-shaped vertical layer that forms along the HFT (see right panel
of Figure \ref{fig_3dfdlcurr} and the 2D cuts shown in Figure \ref{fig_qfac}).
Reconnections in the current layers are found to add to the flux rope
twisted flux threading under the axis, creating
an expanding, low density hot channel in the flux rope above the central
vertical current layer.
When viewed along a line of sight that is roughly aligned with the
channel, it manifests as a central hot void on top of a relatively
dense structure corresponding to the central warped, vertical current layer,
which also extends upward on two sides of the void (see the $J/B$ image at
$t=18.24$ hour in Figure \ref{fig_merievol}), producing
the ``U'' shaped dense shell (or ``horns'') enclosing the void in the modeled
EUV images (see e.g. the $t=12.89$ hour and $t=18.34$ hour panels of the
modeled AIA 171 images in Figures \ref{fig_limbview_evol}). The dense vertical
column and the upward extended horns are essentially the manifestation of the
HFT (see the 2D cut in the top left panel of Figure \ref{fig_qfac}).
Such thermal structures inside the flux rope
are consistent with several
AIA EUV observations of coronal prominence cavities where 
it is found that an elevated, ellipse or tear-shaped central void is
situated on top of the vertical column of dense prominence, and the void
is enclosed by a ``U'' shaped dense shell (sometimes called ``horns'') extending
upward from the prominence column \citep[e.g.][]{regnieretal2011, berger2012}.
The central hot channel produced by the tether cutting reconnections can also
explain the observational result
of a central hot core with low emission measure inside a coronal cavity
by \citet{reevesetal2012}.

\begin{figure}
\epsscale{0.7}
\plotone{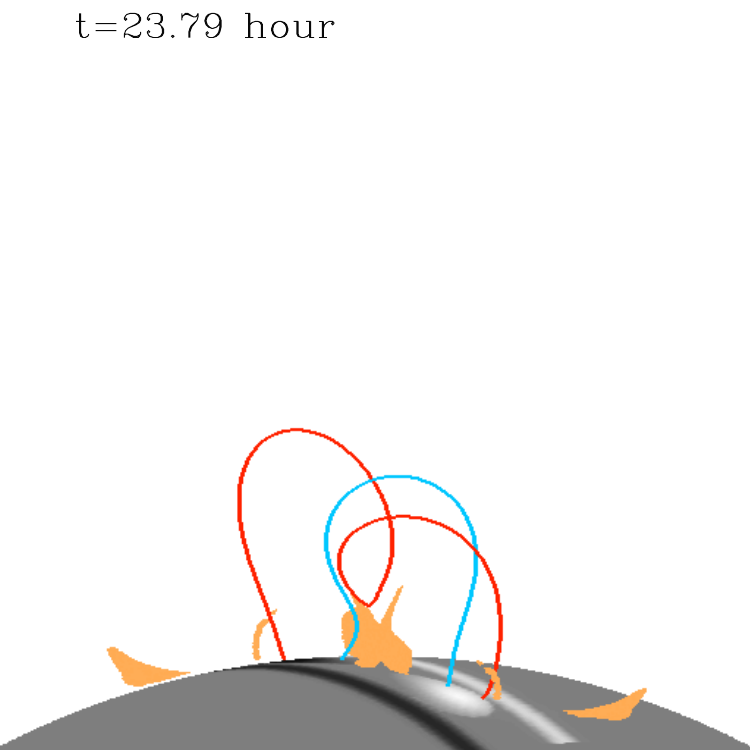}
\caption{A 3D view of the current layers as outlined by the
iso-surface of $J/B$ at a value of $1/(10 dr)$, with dr being the smallest
radial grid size, and two field lines
traced from two points in close proximity on opposite sides of one of the
upward extensions of the central vertical current layer. A movie showing
the 3D rotating view is availabe in the online version.}
\label{fig_horns}
\end{figure}

\begin{figure}
\epsscale{1.}
\plotone{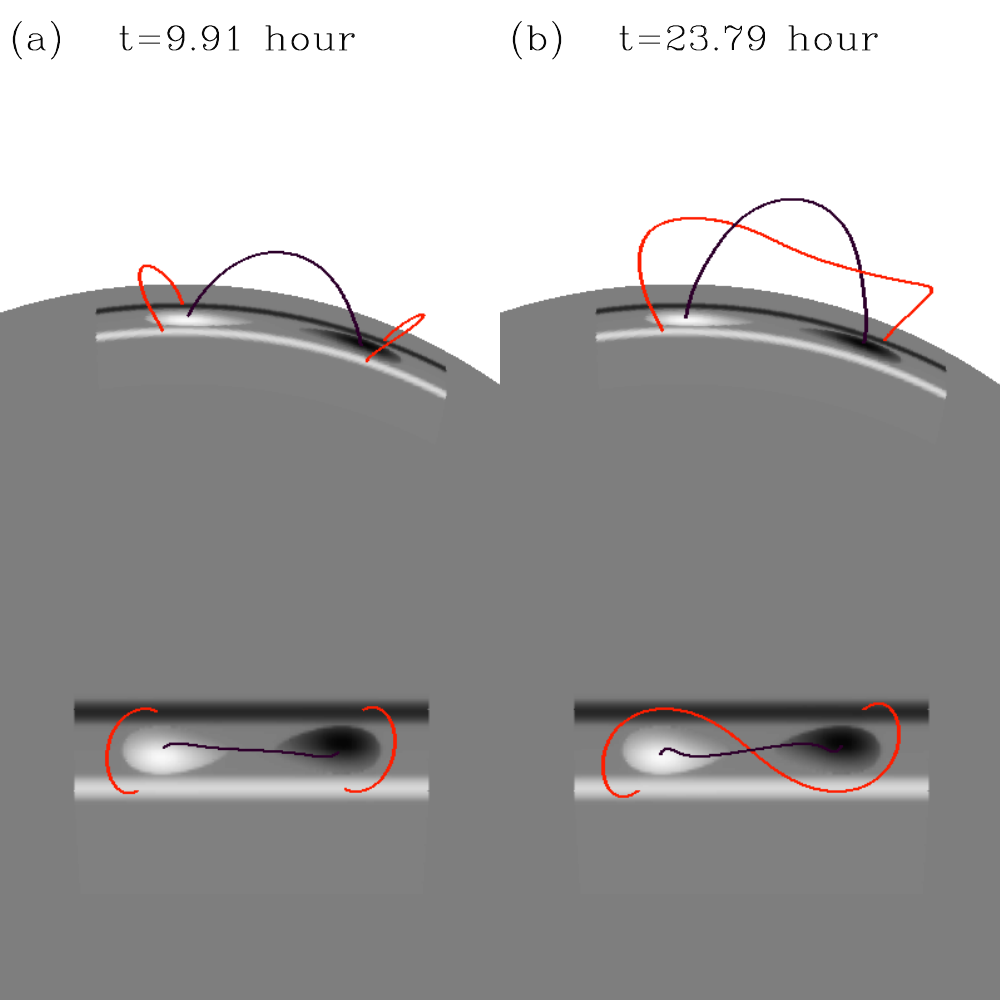}
\caption{3D views of selected field lines from two perspectives. Black field
lines are the tracked axis of the flux rope. The red field lines are field
lines traced from two fixed foot points in the arcade region.  At $t=9.91$
hour, the red field lines are two simple arcade loops.
At $t=23.79$ the red field
line traced from the same two foot points in the arcade region has become a
field line that threads under and twists about the axis of the flux rope, and
it goes goes through the central hot void.}
\label{fig_reconnect}
\end{figure}

\begin{figure}
\epsscale{0.75}
\plotone{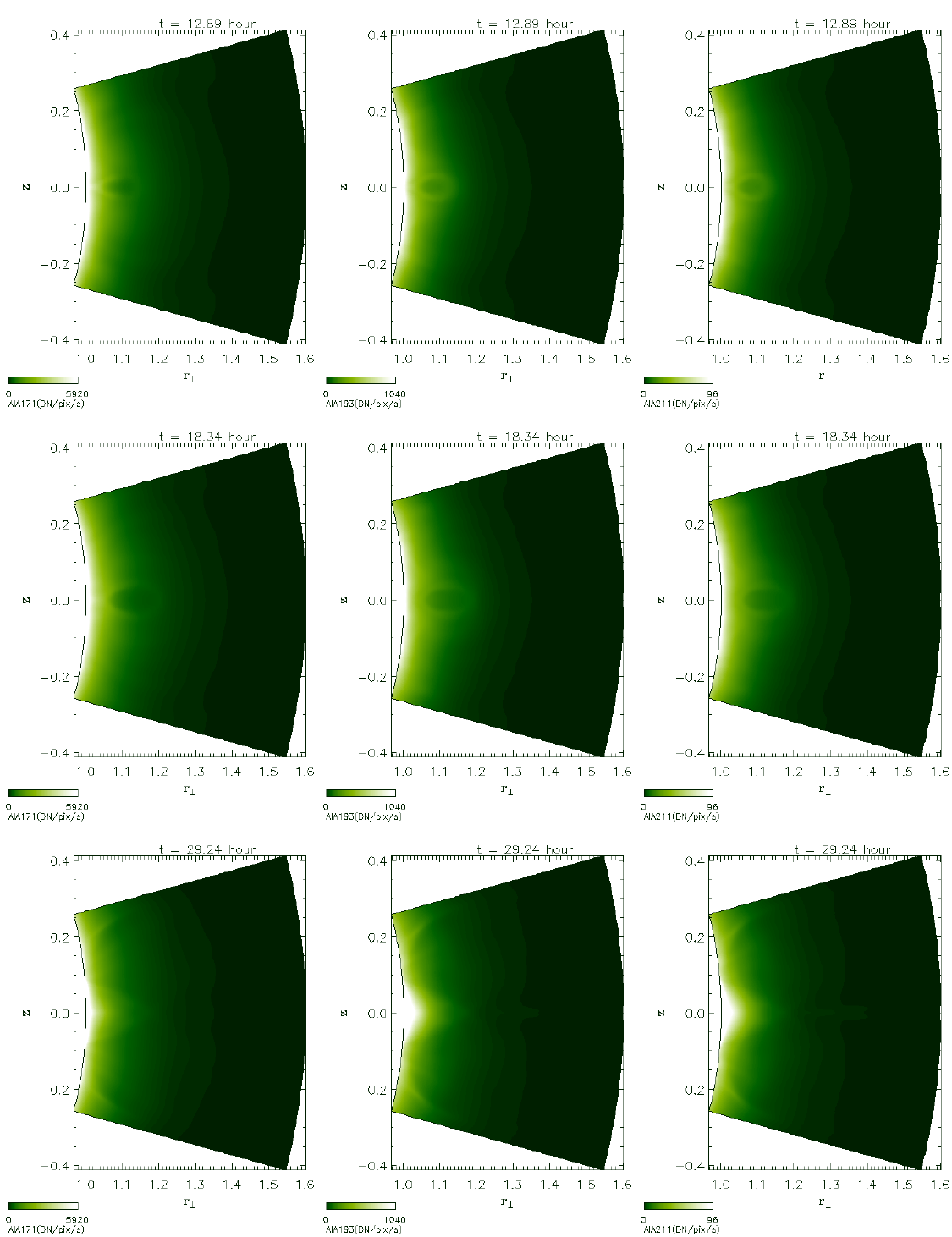}
\caption{Snapshots of the modeled AIA intensity images at 171 {\AA},
193 {\AA}, and 211 {\AA} channels at 3 time instances, showing the expansion
and rise of the central void during the quasi-static phase, and its final
ejection and the post eruption brightening.  A movie of this evolution is
available in the online version of the paper.}
\label{fig_limbview_evol}
\end{figure}

Although our numerical simulation does not include radiative cooling and thus
cannot model the process of prominence formation, it is likely that
the prominence plasma is co-spatial with the dense current layer
\citep[e.g.][]{vanb_cranmer2010, lowetal2012}.
Our MHD simulation also does not model explicitly the background coronal
heating but instead simply assume a high temperature coronal plasma with
an adiabatic
index of $\gamma=1.1$, without explicitly incorporating the coronal heating
and radiative cooling processes. However our model does take into account
localized heating due to the formation and dissipation of current sheets in
the corona and the redistribution of the heat by field aligned conduction
in the corona.  The lower boundary of our simulation domain
is placed at the coronal base instead of the chromosphere, and after the
emergence is stopped, the velocity field is set to zero on the lower boundary
and no more mass flux is allowed through it into the coronal domain.
This precludes the important effect of chromospheric evaporation due to
conductive heat flux coming down from the corona, which may significantly
affect the density of the heated coronal flux tubes. Thus there are
great uncertainties in the density and temperature obtained due to our still
highly simplified treatment of the thermodynamics, especially for the heated,
twisted field lines in the hot channel that form as a result of the tether
cutting reconnections. One main question would be whether the heated field
in the hot channel would remain a void (i.e. of lower density) in contrast
to the surrounding as is in our model when the effect of chromospheric
evaporation produced by the downward heat conduction is taken into account.
In regard to this question, one consideration is that the reconnected field
lines in the hot channel
are generally long field lines (see e.g. the red field lines in panel
\ref{fig_reconnect}(b) and Figure \ref{fig_horns}).  The scaling law derived
based on the static, thermal equilibrium coronal loop model of
\citet{rosneretal1978} gives
$T_{\rm max} \sim 1.4 \times 10^3 (p L)^{1/3}$, where $T_{\rm max}$,
$p$ and $L$ denote respectively the maximum loop temperature,
the loop pressure, and the loop length.
This would mean that the plasma density of the loop $\rho \propto
{T_{\rm max}}^2 L^{-1}$.
Thus even though the field lines in the hot core are heated to
a mildly increased $T_{\rm max}$: to roughly
1.2 MK from the 1 MK original coronal temperature,
these field lines are on the other hand considerably longer in $L$ compared
to the field lines immediately outside the
hot core: for example, in Figure \ref{fig_horns}, the red field line inside
the central void is about 1.9 times the blue field line immediately outside of
the horn.  Therefore the above scaling law of static thermal equilibrium
coronal loops would indicate that the effect of the longer length outweighs
the effect of the high temperature, and the resulting density inside the
central hot channel remains smaller compared to the surrounding, even with
the effect of the chromospheric evaporation included.

Our 3D MHD simulations with the
simplified thermodynamics is an initial step to understand
{\it qualitiatively} the
immediate thermal effect of the formation and dissipation of current sheets 
along the HFT in the corona. Our results
suggest that the observed feature of prominence with upward extending
``horns'' enclosing an elevated central void seen in AIA 171 observations of
coronal cavities is a signature of the
development of the HFT topology and the associated
tether-cutting reconnections in the current sheets that form along the HFT.
It is likely that the current sheet forming along the HFT remains the region
of the highest density because of the significant local compression by
the Lorentz force even when more realistic treatment of the thermodynamics
are included, and our results about the thermal sigature, i.e. relatively
hot, low density central void on top of and surrounded by the higher density
current sheet formed along the HFT, remain qualitatively true.
Of course 3D MHD simulations with a full treatment of the thermal dynamics
that incorporates explicit coronal heating, radiative cooling, as well as
field aligned thermal conduction, and with
the lower boundary set at the chromosphere, are needed
to confirm the findings here and provide quantitative determination of
the density and temperature of the coronal plasma. As has been found in
1D hydrodynamic simulations with a full treatment of the thermodynamics 
along long dipped coronal loops \citep[e.g.][]{karpen_antiochos2008},
complex, thermal nonequilibrium formation
and dynamic evolution of prominence condensations can develop along the
coronal loops depending on the spatial and temporal properties of the heating
profiles prescribed.

Here we have presented the MHD simulation of the quasi-static evolution and
onset of eruption of a particular coronal flux rope configuration.  However
the results presented here of the development of the HFT topology, and
the associated current sheet formation and tether cutting
reconnections along the HFT are quite general results for the evolution of
line-tied coronal flux ropes as has been demonstrated by several previous
MHD simulations as well as NLFFF models of sigmoid active regions based
on observations \citep[e.g.][]{aulanieretal2005, aulanieretal2010, savchevaetal2012a, savchevaetal2012b}.
These models suggest that the development and the elevation of the HFT
may be an indication of the extent to which the tether-cutting reconnections
have progressed and hence the readiness of the flux rope to erupt.

\acknowledgements
I thank B.C. Low for reading and helpful comments on the manuscript.
I thank the anonymous referee for critical review of the paper.
NCAR is sponsored by the National Science Foundation.
This work is supported by the NASA LWS TR\&T grant NNX09AJ89G
to NCAR.
The numerical simulations were carried out on the Pleiades
supercomputer at the NASA Advanced Supercomputing Division under
project GID s0969.

\clearpage

\clearpage

\clearpage

\clearpage

\clearpage

\clearpage

\clearpage

\clearpage

\clearpage

\clearpage

\clearpage

\clearpage

\end{document}